\documentclass[12pt]{article}
\usepackage{times}
\usepackage{geometry}
\usepackage{natbib}
\usepackage[utf8]{inputenc}
\usepackage{enumitem,amssymb}
\usepackage{ragged2e}
\newlist{thematic}{itemize}{8}
\setlist[thematic]{label=$\square$}
\usepackage{pifont}
\newcommand{\cmark}{\ding{51}}%
\newcommand{\done}{\rlap{$\square$}{\raisebox{2pt}{\large\hspace{1pt}\cmark}}%
\hspace{-2.5pt}}

\usepackage{natbib}
\usepackage{aas_macros}
\usepackage[pdftex]{graphicx}
\usepackage{wrapfig}

\begin{document}
\pagestyle{empty}
\raggedright
\huge
Astro2020 Science White Paper \linebreak

The Critical, Strategic Importance of Adaptive Optics-Assisted Ground-Based Telescopes for the Success of 
Future NASA Exoplanet Direct Imaging Missions \linebreak
\normalsize

\noindent \textbf{Thematic Areas:} \hspace*{55pt} $\boxtimes$ Planetary Systems \hspace*{10pt} $\square$ Star and Planet Formation \hspace*{20pt}\linebreak
$\square$ Formation and Evolution of Compact Objects \hspace*{31pt} $\square$ Cosmology and Fundamental Physics \linebreak
  $\square$  Stars and Stellar Evolution \hspace*{1pt} $\square$ Resolved Stellar Populations and their Environments \hspace*{40pt} \linebreak
  $\square$    Galaxy Evolution   \hspace*{45pt} $\square$             Multi-Messenger Astronomy and Astrophysics \hspace*{65pt} \linebreak
  
\textbf{Principal Author:}

Name: Thayne M. Currie	
 \linebreak						
Institution: NASA-Ames Research Center
 \linebreak
Email: thayne.m.currie@nasa.gov
 \linebreak
Phone: (857) 998-9771
 \linebreak
 
\textbf{Co-Authors}: Ruslan Belikov (NASA-Ames Research Center), Olivier Guyon (Subaru Telescope), N. Jeremy Kasdin (Princeton University), Christian Marois (NRC-Herzberg), Mark S. Marley (NASA-Ames Research Center), Kerri Cahoy (Massachusetts Institute of Technology), Dimitri Mawet (California Institute of Technology), Michael McElwain (NASA-Goddard Spaceflight Center), Eduardo Bendek (NASA-Ames Research Center), Marc J. Kuchner (NASA-Goddard Spaceflight Center), Michael R. Meyer (University of Michigan) 
\linebreak
\linebreak
\textbf{Co-Signers}: S. Mark Ammons (Lawrence Livermore National Laboratory), Julien Girard (Space Telescope Science Institute), Yasuhiro Hasegawa (Jet Propulsion Laboratory), Mercedes Lopez-Morales (Center for Astrophysics | Harvard \& Smithsonian), Wladimir Lyra (California State University-Northridge/Jet Propulsion Laboratory),  Ben Mazin (University of California-Santa Barbara), Bertrand Mennesson (Jet Propulsion Laboratory),  Chris Packham (University of Texas-San Antonio),  Tyler Robinson (Northern Arizona University)
  \linebreak

\textbf{Abstract}: Ground-based telescopes coupled with adaptive optics (AO) have been playing a leading role in exoplanet direct imaging science and technological development for the past two decades and will continue to have an indispensable role for the next decade and beyond.
Over the next decade, extreme AO systems on 8-10m telescopes will 
1) mitigate risk for WFIRST-CGI by identifying numerous planets the mission can spectrally characterize, 2) validate performance requirements and motivate improvements to atmosphere models needed to unambiguously characterize solar system-analogues from space, and 3) mature novel technological innovations useful for space.
Extremely Large Telescopes can deliver the first thermal infrared (10 $\mu m$) images of rocky planets around Sun-like stars and identify biomarkers.   These data provide a future NASA direct imaging flagship mission (i.e. HabEx, LUVOIR) with numerous exo-Earth candidates and critical ancillary information to help clarify whether these planets are habitable.


\pagebreak
\pagestyle{plain}
\setcounter{page}{1}
\justifying
\begin{figure*}
    \centering
    \vspace{-0.1in}
    \includegraphics[scale=0.875,trim = 0mm 2mm 0mm 5mm,clip]{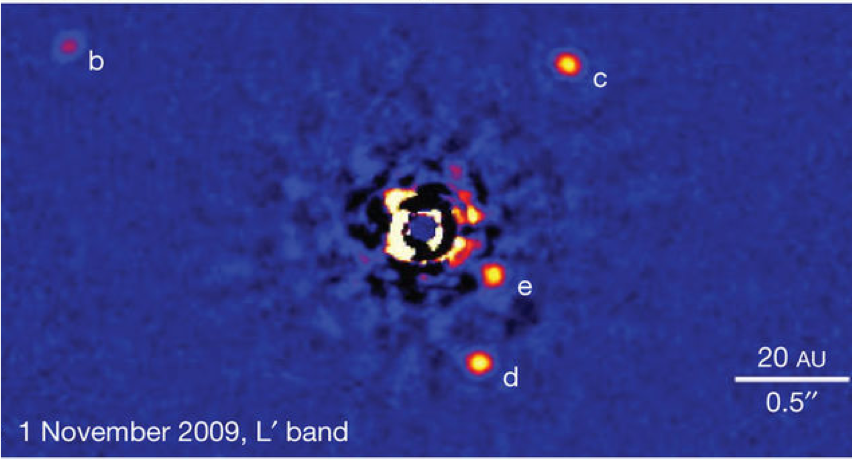}
    \vspace{-0.135in}
    \caption{The first directly-imaged planetary system, HR 8799: discovered with Keck and Gemini \citep{Marois2008,Marois2010}.} 
    \label{fig:hr8799}
    \vspace{-0.165in}
\end{figure*}
\vspace{3mm}
\noindent {\bf 1.~The Ground’s Long-Standing, Critical Role in Advancing Exoplanet Direct Imaging}
\vspace{1mm}

Since the first nearby extrasolar planets were discovered from indirect detection methods \citep{Mayor1995}, NASA has shown a strong interest in someday directly imaging and spectrally characterizing solar system-like planets, including an Earth twin around a Sun-like star \citep[e.g. the Terrestrial Planet Finder mission,][]{Beichman1998}. 
 
Yet nearly two decades later, ground-based telescopes utilizing adaptive optics (AO) have accounted for nearly \underline{all} of the 15 to 20 exoplanets and candidates imaged thus far \citep[e.g. Fig.1;][]{Marois2008,Bowler2016}.  Follow-up photometry and spectra of these first discoveries yielded crucial insights into young super-Jovian planets’ atmospheric properties: clouds, carbon chemistry, and surface gravity \citep[e.g.][]{Currie2011,Marley2012}.   Some recent exoplanet direct imaging discoveries – enabled by extreme AO achieving deeper contrasts than facility systems – have revealed cooler, lower-mass planets with far different spectra \citep{Macintosh2015}. 

Ground-based telescopes have also matured key technologies and methods forming the backbone of future NASA direct imaging missions.   The integral field spectrograph,  first developed/matured on the ground and coupled to extreme AO systems \citep[e.g.][]{McElwain2007,Larkin2014},  is now considered to be a key component of any future exo-Earth imaging system.  
The ground has demonstrated advanced coronagraph designs like the vector vortex \citep{Mawet2010}, which is being considered for HabEx.  Sophisticated post-processing and spectral extraction methods were developed from analyzing ground-based data and demonstrated to improve planet imaging capabilities from space \citep{Lafreniere2007,Marois2010,Soummer2012} .   Ground-based data clarified how speckle noise statistics relate to spectrophotometric uncertainties and affect planet detection limits \citep{Marois2008b,Mawet2014,Greco2016}.


As described below, {\bf the ground can play an indispensable role in ensuring the success of NASA exoplanet direct imaging} missions -- through both science and technological insights -- {\bf for the next decade} of extreme AO on current telescopes \textbf{and} with ELTs \textbf{for decades to come}.

\begin{figure*}
    \centering
    \includegraphics[scale=1.165,trim=0mm 0mm 6mm 0mm,clip]{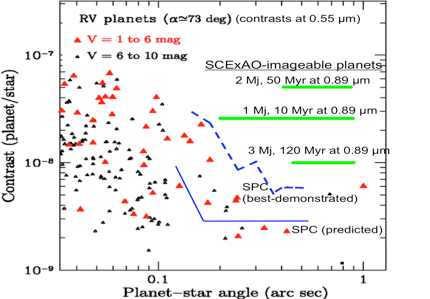}
    \includegraphics[scale=1.025,trim=0mm 0mm 0mm 7mm, clip]{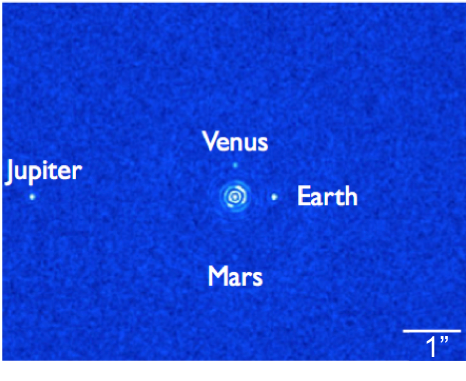}
    \vspace{-0.175in}
    \caption{(Left) Even if the WFIRST-CGI Technology Demonstrator does not quite reach to reach contrasts necessary to image mature Jupiter twins in reflected light ($\sim$ 10$^{-9}$), 
    over the next decade ground-based extreme AO systems utilizing focal-plane wavefront control (e.g. Subaru/SCExAO) could provide WFIRST-CGI with numerous  younger, cool self-luminous jovian planets to spectrally characterize.  Discovery space from the ground is noted by green bars; blue lines denote the predicted and best-demonstrated (in a lab) WFIRST-CGI contrasts. 
    (Right) Simulated  10 $\mu m$ image of the inner solar system at 1.3 pc (credit: C. Marois).  At 10 $\mu$m, ELTs can image Earth-like planets around $\sim$ 10 AFGK stars, which could be followed up with a NASA mission like HabEx or LUVOIR.}
    \label{fig:groundscience}
    \vspace{-0.2in}
\end{figure*}
\vspace{3mm}
\noindent {\bf 2. The Next Decade of Ground-Based Extreme AO Systems: ~Advancing the Scientific Potential of and Mitigating Risk for WFIRST-CGI (2020-2029)}
\vspace{1mm}

The first generation of extreme AO systems demonstrated the capability to create a sub-arcsecond dark hole via high-order deformable mirrors (DM) driven by advanced wavefront sensors \citep[e.g.][]{Esposito2011,Sauvage2016}.  Coupled with advanced coronagraphy, they achieved contrasts of 10$^{-5}$--10$^{-6}$ at 0.2"--0.5".

Over the next decade, extreme AO systems on current  ground-based class telescopes will build upon this foundation and break new ground.   Instruments such as Subaru/SCExAO, GPI-2.0, and MagAO-X can/will utilize 
focal-plane wavefront control techniques first developed in the lab and now tested on sky \citep[e.g.][]{Martinache2014,Bottom2016,Matthews2017}.   They can 
be further equipped with ultra-low noise detectors (e.g. MKIDS) to drive focal-plane wavefront control at very high speeds \citep[e.g.][]{Mazin2012}.   As a result, by the latter half of the 2020s, at least some extreme AO systems should deliver much deeper contrasts than currently achieved, approaching 10$^{-7}$ or possibly lower at small angles in the near-infrared (near-IR).

These new extreme AO capabilities in turn open up new exoplanet discovery and characterization space relevant for WFIRST-CGI, which should be operational within a decade.  For example,  
a factor of 10 improvement in contrast over GPI/SPHERE at 0.1"--0.5" ($\sim$ 10$^{-6}$--10$^{-7}$) could allow the detection of 1--3 $M_{\rm J}$, 10-120 Myr-old planets orbiting at Jupiter-like separations for stars in the nearest moving groups. In the red optical, these planets should have contrasts just 20—30 times fainter (10$^{-7}$--10$^{-8}$) and thus should be easily recoverable with WFIRST-CGI (Marley et al. 2019, in prep.; Fig. \ref{fig:groundscience}a).   They probe a phase space poorly explored in planetary atmospheres: 375–500 K objects near the T to Y dwarf transition but at much lower gravity than field brown dwarfs.   Like the first directly-imaged exoplanets, these newly-characterized exoplanets will challenge existing models and lead to a richer understanding of key atmospheric properties like clouds and chemistry but at temperatures and masses more characteristic of solar system planets \citep{Marley2019}. 

Hosts for planets discovered with extreme AO systems over the next decade will also include early type stars (e.g. Vega, Fomalhaut, Altair) that provide WFIRST-CGI with a much more photon-rich environment for wavefront control and whose planets are inaccessible to precision radial-velocity detection.   Dedicated WFIRST-CGI precursor surveys on 8–10 m class telescopes operating with extreme AO could identify a large sample of WFIRST-accessible self-luminous planets.  If $\sim$ 10$^{-8}$ contrast is reached from leading/upcoming extreme AO systems on current telescopes, some mature, $\sim$ 1 au-separation jovian planets in reflected light are reachable\footnote{Planets suitable for WFIRST-CGI follow-up could also be imageable with first-light 3–5 $\mu$m instruments on ELTs.}.    

\textbf{Thus, even if WFIRST-CGI fails to deliver the contrasts needed to image mature reflected-light Jupiter twins (10$^{-9}$), the next decade of ground-based extreme AO systems could provide the mission many younger, brighter Jupiters to demonstrate and advance precise atmosphere parameter estimation}.   

\vspace{3mm}
\noindent {\bf 3.~Direct Imaging Observations of Exo-Earths with Extremely Large Telescopes: Crucial Preparatory/Complementary Data for a NASA Flagship Direct Imaging Mission (2025+)}
\vspace{1mm}

While space may be needed to directly detect an Earth twin around a Sun-like star in reflected light, ELTs will be able to image exo-Earths in thermal emission at 10 $\mu$m (Fig \ref{fig:groundscience}b).  Multiple ELTs are considering first-generation instruments with these capabilities: i.e. E-ELT/METIS and TMT/MICHI.    These instruments couple low-emissivity AO delivering exceptionally high-Strehl corrections with advanced coronagraphy to yield contrasts at r $<$ 1" separations of $\sim$10$^{-7}$ in several hours of integration time.  At 10 $\mu m$, the habitable zone for Sun-like stars lies beyond 2.5--3 $\lambda$/D for ELTs out to 5--6 pc.  Direct imaging of rocky,  Earth-radius planets receiving the same insolation as the Earth is possible for ELTs around $\sim$ 10 nearby AFGK stars, including $\epsilon$ Eridani, $\tau$ Ceti, and Procyon; ozone may be detectable with focused follow-up observations \citep{LopezMorales2019}.   The 10 $\mu$m discovery space for Super-Earths receiving Mars-like insolation (50\% larger separation) or warm, Neptune-sized planets is even larger.

\textbf{Direct imaging of rocky planets with ELTs}, especially in the thermal infrared, also \textbf{substantially increases the science gain of NASA direct imaging missions}.    Disentangling the effect of a planet’s radius from its albedo based on reflected-light spectra alone is extremely challenging.  However, thermal infrared data helps to constrain the equilibrium temperature of rocky planets and in turn the planet radius.  Multi-epoch 10 $\mu$m imaging data will help identify optimal times for reflected-light space observations.  
For nearby low-mass stars, ground-based reflected-light imaging \citep[e.g. \textit{The Planetary Systems Imager} on TMT;][]{Guyon2018} can identify exo-Earth candidates showing evidence for biomarkers in the near-IR (e.g. O$_{\rm 2}$ at 1.27 $\mu$m) \citep{LopezMorales2019}.  Some of these could be reimaged in the optical with large-aperture NASA direct imaging missions probing different biomarkers, yielding more robust assessments about habitability.  

In summary, \textbf{direct imaging programs with ELTs can identify the best exo-Earth candidates for follow-up with HabEx/LUVOIR, increasing the missions' yields for and improving their characterization of habitable zone, rocky planets}.   ELTs may be especially critical for missions employing a starshade for spectroscopic follow-up, which will require well-vetted targets. 

\begin{figure*}
    \centering
    \includegraphics{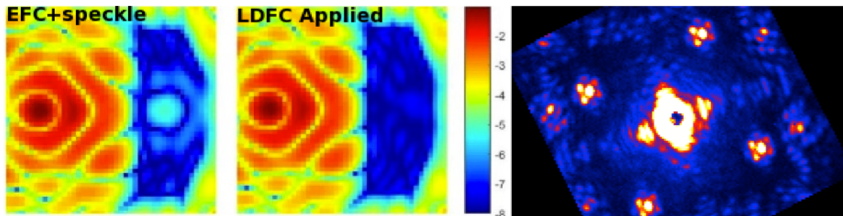}
    \vspace{-0.2in}
    \caption{Novel technologies that can be matured on the ground and applicable to future NASA missions.  Simulation demonstrating Linear Dark Field Control (LDFC) \citep{Miller2017}.   (Right) SCExAO/CHARIS data using the shaped-pupil coronagraph, tests for which assessed the design's sensitivity to low-order aberrations.}
    \label{fig:techdev}
    \vspace{-0.15in}
\end{figure*}
\vspace{3mm}
\noindent {\bf 3.~NASA-Applicable Technical Assessments and Technology Developments from Ground-Based Extreme AO}
\vspace{1mm}

Future NASA direct imaging platforms from WFIRST-CGI to HabEx/LUVOIR are planned to yield contrasts below 10$^{-8}$ to nearly 10$^{-11}$ at small angles, sufficient to detect and spectrally characterize mature Jupiter twins to Earth-sized, habitable zone rocky planets.   Laboratory investments in wavefront control and coronagraphy and detailed simulations to optimize performance, recommended by the 2010 Decadal Survey, have significantly advanced these goals  \citep[e.g.,][]{Cady2016,Mazoyer2016,Leboulleux2017}.  Due to atmospheric turbulence, ground-based extreme AO faces unique technical challenges and greater performance limitations.
Reaching contrasts much deeper than 10$^{-8}$ on ELTs on the even the most laminar sites (i.e. Maunakea) will be exceedingly challenging \citep{Stapelfeldt2006}, 
although new wavefront sensing and control advances 
will push limits \citep[e.g.][]{MalesGuyon2017}.   While  current and future \textbf{ground-based telescopes} will operate at much brighter contrast limits than upcoming NASA direct imaging missions, the ground \textbf{can provide NASA-applicable technical assessments and help mature novel technological advances applicable to future NASA direct imaging missions}.

For instance, ground-based extreme AO may provide insights about spectral retrieval in the presence speckle noise useful for upcoming NASA missions.   At contrasts achieved by current ground-based extreme AO systems (10$^{-4}$--10$^{-6}$) relatively achromatic phase errors dominate the speckle noise budget.   At contrasts closer to WFIRST-CGI’s performance (10$^{-7}$--10$^{-9}$) chromatic errors should begin to dominate \citep[e.g.][]{Bailey2018}. This change in regime affects post-processing and the spectral covariance of integral field spectrograph data, which impacts the signal-to-noise ratio (SNR) needed for a given atmospheric characterization goal \citep{Greco2016}.  Future ground-based extreme AO systems, especially those on ELTs, 
approaching 10$^{-7}$--10$^{-8}$ contrast at several $\lambda$/D will begin to sample the regime over which this transition occurs.    
Analysis of \textbf{extreme AO data over the next decade and with ELTs can help establish ``best practices" for} spectral retrieval, guiding estimates for the
required SNR for atmosphere characterization goals in the extremely faint speckle regime in which \textbf{future NASA missions} will operate.

\textbf{The ground also provides a nimble path to mature novel technological innovations in wavefront control methods potentially applicable to a future NASA mission}.   For example, Subaru/SCExAO and MagAO-X will help mature Linear Dark Field Control (LDFC), which utilizes the linear response of the region outside a dark hole (i.e. the ``bright field") dug by focal-plane wavefront control methods like electric field conjugation (EFC) to correct wavefront perturbations that affect both the bright field and dark field \citep[Fig \ref{fig:techdev}a][]{Miller2017}.    While EFC uses DM probes to update the estimate of the electric field which perturb the science exposure, LDFC freezes the DH state initially achieved with EFC, potentially allowing far greater observing efficiency, a more static dark hole speckle halo, and thus deeper contrasts.   Ground-based tests will provide a first assessment of the dark hole stability with LDFC vs. current leading methods and its practical performance advantage when incorporated within the context of a high-contrast imaging system.   Multi-star wavefront control is likewise being matured on ground-based extreme AO systems (i.e. SCExAO) and may be important for upcoming NASA missions \citep{Sirbu2017}, since multiple key direct imaging targets are binary systems (e.g $\alpha$ Cen).   

New wavefront control and coronagraphy tested within the context of a complete extreme AO system may reveal information hidden from targeted laboratory experiments.   For example, tests of the shaped-pupil coronagraph (SPC) with extreme AO assessed its sensitivity to low-order aberrations for Strehl ratios relevant for both extreme AO and future NASA missions \citep[Fig. \ref{fig:techdev}b;][]{Currie2018b}.  While qualitatively the SPC performed as expected (low sensitivity to low-order modes), the on-sky data revealed that the relative advantage of the SPC over other designs may be larger than expected from laboratory tests alone, as assessed by changes in the intensity profile of the dark hole as the fidelity of the AO system correction for low-order modes was varied.
 
\vspace{3mm}
\noindent {\bf 4.~Recommendations}
\vspace{1mm}

The 2010 Decadal survey noted the importance of strategic ground-based efforts to constrain exoplanet demographics (i.e. with indirect methods) and typical levels of exozodiacal light (e.g. with the LBTI/HOSTS program).  It also included technology development for  future direct imaging missions in its top-ranked medium scale priority.
Last year, the National Academy of Sciences {\it Exoplanet Science Strategy Consensus Study Report} (NAS-ESS) recommended both an investment in ELTs and a future NASA flagship mission to image exo-Earths.

As argued here, these two NAS-ESS recommendations are symbiotic: \textbf{ground-based direct imaging}, both now and later with ELTs, through both science and technology development, \textbf{is a key, strategic investment to help ensure the success of future NASA direct imaging missions} from WFIRST-CGI to HabEx or LUVOIR.   To maximize the return that the ground can give for NASA direct imaging missions we make the following recommendations:
\begin{itemize}
\setlength\itemsep{0.1em}
  \item[1-]  \textbf{Ground-based direct imaging science programs with extreme AO on 8-10m class telescopes and later on ELTs should be explicitly considered as critical, necessary support for NASA’s direct imaging-focused missions}.   Precision radial-velocity surveys have been emphasized as key ground support for future NASA-missions like WFIRST-CGI \citep{Dressing2019} as they yield mature jovian planets the mission can detect and characterize and can constrain masses.  Similarly, \textbf{extreme AO on ground-based telescopes} over the next decade \textbf{can provide key WFIRST-CGI support}.   Critically, extreme AO programs can identify key targets that do not require 10$^{-9}$ contrast for detection and yield supporting measurements crucial for atmospheric characterization with WFIRST-CGI.
Furthermore,  10 $\mu$m (and possibly reflected-light) \textbf{direct imaging with ELTs} likewise precedes and \textbf{can provide critical information needed to identify habitable, Earth-like planets with HabEx/LUVOIR}.  

\item[2-] \textbf{NASA should take an explicit interest in and collaborate on ground-based technology developments} -- focal-plane wavefront control, coronagraphy, advanced post-processing/spectral retrieval methods, and the practical gains achieved by integrating these components, etc.   If the past is any guide, \textbf{these efforts over the next decade can generate key, novel advances applicable to NASA direct imaging missions}.   
 \end{itemize}
 
 To optimize the ground's utility for NASA missions, one option is to support more formal, direct partnering between NASA laboratories and observatories, an extension of the NAS-ESS's recommendation of individual investigator programs.   Doing so allows for a cross-pollination of ideas benefiting both NASA and ground-based observatories in the near term and helps the next generation of instrument scientists hone their skills so that they can advance NASA direct imaging missions in the long term. 

\pagebreak
\setlength{\bibsep}{0.0pt}

\end{document}